\documentclass [preprint]{aastex}
\usepackage{longtable}

\shorttitle{Highly Variable Objects in the P-Q Survey}
\shortauthors{Bauer et al.}

\begin{document}

\title{Highly Variable Objects in the Palomar-QUEST Survey: \\A Blazar Search using Optical Variability}

\author{Anne Bauer\altaffilmark{1,2}, Charles Baltay\altaffilmark{2}, Paolo Coppi\altaffilmark{2}, Ciro Donalek\altaffilmark{3}, Andrew Drake\altaffilmark{3},  S. G. Djorgovski\altaffilmark{3}, Nancy Ellman\altaffilmark{2}, Eilat Glikman\altaffilmark{3,2}, Matthew Graham\altaffilmark{3}, Jonathan Jerke\altaffilmark{2}, Ashish Mahabal\altaffilmark{3}, David Rabinowitz\altaffilmark{2}, Richard Scalzo\altaffilmark{2}, Roy Williams\altaffilmark{3}}
\email{anne.bauer@usm.uni-muenchen.de}
\altaffiltext{1}{Universit\"{a}ts-Sternwarte M\"{u}nchen, Scheinerstr. 1, D-81679 M\"{u}nchen, Germany}
\altaffiltext{2}{Yale University, Department of Physics, P.O. Box 208120, New Haven, CT 06520-8120, USA}
\altaffiltext{3}{California Institute of Technology, Department of Astronomy, MC 105-24, 1200 East California Blvd, Pasadena CA 91125, USA}

\begin{abstract}

We identify 3,113 highly variable objects in 7,200 square degrees of 
the Palomar-QUEST Survey, which each varied by more than 0.4 
magnitudes simultaneously in two broadband optical filters on timescales 
from hours to roughly 3.5 years.  The primary goal of the selection 
is to find blazars by their well-known violent optical variability.  Because 
most known blazars have been found in radio and/or X-ray wavelengths, 
a sample discovered through optical variability may have very different 
selection effects, elucidating the range of behavior possible in these 
systems.  A set of blazars 
selected in this unusual manner will improve our understanding of the 
physics behind this extremely variable and diverse class of AGN.  
The object positions, variability statistics, and color 
information are available using the Palomar-QUEST CasJobs server.  
The time domain is just beginning to be explored over 
large sky areas;  we do not know exactly what a violently 
variable sample will hold.  About 20\% of the sample has been 
classified in the literature;  over 70\% of those objects are known or 
likely AGN.  
The remainder largely consists of a variety of variable stars, including 
a number of RR Lyrae and cataclysmic variables.  

\end{abstract}

\keywords{galaxies: active --- quasars: general --- BL Lacertae objects: general --- catalogs }

\section{Introduction}

Palomar-QUEST (PQ) is one of the first surveys with repeated observations 
over a large fraction of the sky, and has the potential to find many 
rare variables.  We have carried out a search for highly variable objects; 
while it is interesting to see the make-up of such a variability-selected 
list, the search is motivated by the desire to find blazars.  A large 
optically identified set of blazars with selection effects different 
from current samples will help us to understand the range of 
blazar behavior and its underlying mechanisms.

Blazars are active galactic nuclei (AGN) that emit jets and are aligned such 
that the jet axis is within about $15^{\circ}$ of the line of sight.  
The jets are made up of highly energetic photons, electrons, positrons, and 
perhaps also hadrons which emit synchrotron radiation.  This emission, or 
thermal emission from ambient particles, is Compton upscattered by the 
jet particles.  The resulting spectral energy distribution (SED) exhibits 
two large bumps:  a synchrotron peak typically centered in the radio 
and infrared range, and a higher energy peak spanning X-ray or gamma ray 
wavelengths.  The 
jet and its surroundings are inhomogeneous, the flux 
emanating from the shocks depends on the density of the matter involved, 
and the emission is highly beamed in our direction; blazar flux 
is therefore extremely variable in time (see, e.g., measurements by 
\cite{heidt96}; \cite{ciprini03}; \cite{kartaltepe07}).  

Blazars are divided into 
two classes according to their optical spectral properties, in particular by 
how strongly the jet continuum dominates over flux from the accretion disk 
and emission line regions.  BL Lacs have spectra that are featureless, 
where no emission lines 
show through the radiation from the jet.  Flat spectrum radio quasars (FSRQs) 
instead show visible broad optical emission lines.  

The central frequencies of blazars' two SED peaks appear to be correlated.  
The frequencies of the peaks depend 
on the highest energy of the electrons in the jet or the strength of the jet's 
magnetic fields, depending on the jet's emission mechanism 
(\cite{fossati98}).  The 
more energetic the jet particles or strong the magnetic field, the higher the 
frequencies.  It is not clear, however, how these parameters relate to other 
properties of the blazar.  
BL Lacs are divided into two categories according to their peak frequencies:  
HBLs (high peaked BL Lacs) have synchrotron peak frequencies in the UV 
to soft X-ray bands, while LBLs (low peaked
 BL Lacs) have the synchrotron peak in the radio and IR.  The HBL/LBL 
dichotomy was originally used to describe BL Lacs only, although FSRQs can 
be discussed in these terms as well.  Most FSRQs have SEDs similar to LBLs, 
although some have SEDs more like HBLs (\cite{padovani07}).  HBLs and LBLs 
show differences beyond their SED peak frequencies.  For example, 
LBLs are usually 
more luminous than HBLs, and in LBLs the high frequency Compton peak 
dominates while in HBLs the low frequency synchrotron peak tends 
to dominate.  

Currently known blazars have almost entirely been found either by their bright 
radio or X-ray emission, or a combination of both.  HBLs have traditionally 
been found using X-ray data, LBLs using radio data.  Selection effects 
may significantly influence the types of blazars discovered to date.  
A sample of optically selected blazars 
may include many with atypical peak frequencies, 
perhaps in the continuum between the LBL and HBL blazars.  
Transitional objects with some features of both quasars and blazars may 
also be found (e.g., 3C 273; see \cite{kataoka02}; \cite{perlman08}).  
\cite{collinge05} and \cite{londish07} have selected candidate BL Lacs 
by looking for featureless optical spectra among color-selected quasar 
candidates in the SDSS and the 2dF and 6dF surveys, 
respectively.  The samples need further follow-up to be 
securely identified;  however, they appear to have roughly similar SEDs to 
known blazars (particularly HBLs), although they are perhaps 
less radio loud.  A set of optical variability-selected blazars, 
with selection biases different from the current samples, 
would place important constraints on the nature of the different blazar 
types.  

\section{The Palomar-QUEST Survey}

The Palomar-QUEST Survey has observed 15,000 square degrees of sky repeatedly 
using seven optical filters.  
It is a multipurpose project and is currently being used to study 
type I quasars (\cite{quasar_paper}), blazars (\cite{blazar_paper}), 
type Ia supernovae (\cite{snfactory}), transients in real-time 
(\cite{george}), brown dwarfs
(\cite{cathy}; \cite{cathy2}), and solar system objects (\cite{planet}),
among other phenomena.

The data were 
taken over a span of about 3.5 years using the QUEST2 Large Area Camera on 
the Samuel Oschin Schmidt 48'' telescope at Palomar Observatory.  The camera, 
built specifically for this survey, has a $3.6^{\circ} \times 4.6^{\circ}$ 
field of view populated with 112 CCDs.   The CCDs are arranged in 4 rows by 28 
columns such that a different filter can be placed over each row.  
In driftscan mode, an 8 hour scan will yield about 500 square degrees of data 
with successive $\sim140$ second exposures in the four filters.  Two filter 
sets were used during the survey: Johnson UBRI and Gunn r'i'z'z'.  Details of 
the camera are presented in \cite{camera_paper}.

Palomar-QUEST's area coverage includes declinations from -25 to 
+25 degrees, and all right ascensions further 
than 15$^{\circ}$ from the galactic plane.  The amount of time between 
observations over the same coordinates ranges from several hours to the length 
of the survey; typically an area is covered twice in one lunation and revisited 
each year.  On average, there are roughly 5 observations of the entire survey 
area in each filter set.  Certain regions have been examined more often 
for either calibration reasons or specialized analyses, yielding up to 25 
observations for a small fraction of the area.

The survey comprises about 15 terabytes of image data, which have been 
processed with multiple pipelines.  The Yale pipeline, used in this 
work, performs object detection, astrometry, 
photometry, and associates detections of the same object across the four 
rows of chips (or, equivalently, the four color filters).  Photometric 
calibrations correct for sensitivity variations within each chip and between 
different chips, including the effects of non-linearities in some CCDs.  
The survey has amassed its large dataset by observing in non-photometric 
conditions.  Extinction corrections calibrate the data to the level of the 
best PQ scan over each square degree, which is assumed to be photometric.  
This processing software and standard calibration of the data 
are described in detail in \cite{software_paper}.

An additional, relative calibration is done before using the data for variability 
work.  The goal of the relative calibration is not to determine most accurately an 
object's magnitude, but for each individual measurement of an 
object to be as consistent with each other measurement as possible.  The 
extinction correction is therefore eliminated in favor of a routine that 
calibrates the data to the most phometrically stable scan rather than 
the one in which the stars appear brightest.  An additional correction is made 
to account for spatially dependent variations in the data due to effects like 
scattered light in the camera.  Finally, strict quality cuts are implemented 
to eliminate calibration tails due to noisy data, poor measurements of objects that 
do not conform to the PSF profile, or rare catastrophic failures of the 
processing software.  
In order to generate as many comparable measurements of an object as possible, 
we calibrate the Johnson R and SDSS r' measurements, and similarly the Johnson I and 
SDSS i' measurements, together to obtain combined data which we call Rr and 
Ii bandpasses.  The result of the relative calibration is a systematic error 
of 0.7\% for the Rr data, and 1.3\% for the Ii' data.  The relative calibration 
procedure is discussed and evaluated in detail in \cite{quasar_paper}.  The 
relatively calibrated Palomar-QUEST data are used for the variability 
work in this paper.  The standard calibration, in which each of the Johnson 
and SDSS filters is calibrated to be photometric, is used when studying average 
colors of the objects.

\section{Selection Criteria}

\subsection{Variability Cut}

Blazars exhibit high amplitude variability on all timescales probed 
by the Palomar-QUEST Survey.  Their appearence in the survey is examined 
in detail in \cite{blazar_paper}, which examines the ensemble optical 
variability of 276 FSRQs and 86 BL Lacs taken from a variety of sources: 
\cite{1jansky}, \cite{hewitt93}, \cite{collinge05}, \cite{donato05}, 
\cite{gamma1}, \cite{vcv}, \cite{massaro07}, 
\cite{roxa}, and \cite{cgrabs}.  The structure function, 
variability amplitudes over different time lag ranges, and duty cycles 
of these objects are studied in order to characterize the behavior of 
blazars in large-scale variability surveys such as Palomar-QUEST.  
Because blazars vary dramatically 
on all measured timescales, we do not impose frequency-dependent 
selection criteria when compiling the sample.  Instead, we simply 
select all objects that vary by more than 0.4 magnitudes between any 
two epochs of observation.  
A separate search is underway at Caltech for objects that are 
highly variable on short timescales ($\sim$1 hour to a few days);  
this work will be presented in a later paper.  

Our specific cut of 0.4 magnitudes is motivated by the comparisons made 
between type I quasars and blazars in 
\cite{blazar_paper}.  Figures 3 and 4 in that work show the fluctuation 
amplitudes of these AGN classes as seen in the Palomar-QUEST Survey over 
a variety of timescales.  At all timescales measured, quasar 
fluctuations are infrequent at amplitudes greater than a few tenths of 
a magnitude, while the blazar sample includes a significant tail out to 
larger amplitudes.  In order to select jet-based blazar fluctuations 
rather than quasar-like behavior due to accretion disk flares, 
we impose a variability cut at 0.4 magnitudes.
To exclude cases of calibration error, we insist that both the 
Rr and Ii data must show a magnitude change of at least 0.4 magnitudes 
in the same direction between the two epochs.

\begin{figure}
\begin{center}
\plotone{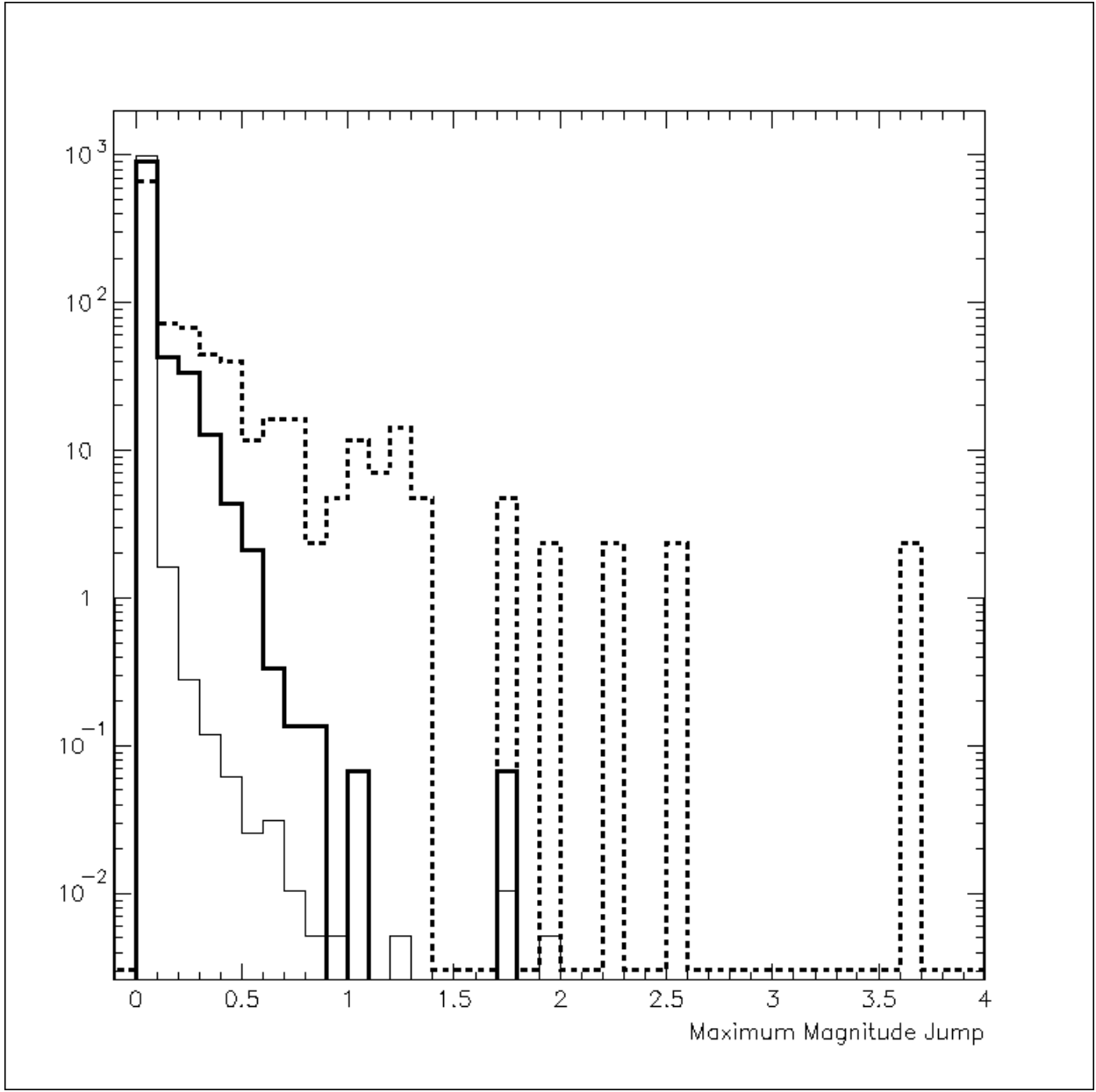}
\end{center}
\caption{Maximum magnitude jump for different object types: a random sample of QUEST objects (thin solid line), quasars (thick solid line), and blazars (dashed line).  The jump must be significant to 3$\sigma$ and seen in both Rr and Ii bands.}
\label{max_jump_fig}
\end{figure}

Figure \ref{max_jump_fig} describes the variability of several types of 
objects in the 
Palomar-QUEST data.  The X axis is the maximum magnitude 
jump seen by each object between any two epochs.  The jump must be significant 
to 3 sigma in both the Rr and Ii bands in order to be included in the histogram.  
The thin solid line describes the variability seen in a random 
subset of 194,000 relatively calibrated PQ objects.  The thick solid 
line denotes 14,800 type I quasars spectroscopically identified by 
the Sloan Digital Sky Survey (SDSS) (\cite{sdss7}).  
The dashed line shows the behavior seen in blazars, using a sample 
of 425 collected in \cite{blazar_paper}.  (We do not insist on redshift 
measurements for these objects, allowing us to use more blazars from 
the surveys sampled by \cite{blazar_paper} than were used in the 
paper's main analysis.)  
The three samples' histograms are each normalized to a total of 1,000 
objects in order to be easily comparable.  

Because we only see roughly 4 epochs for each object, the lightcurves 
for each object are sparsely sampled.  There could easily 
be variability that we do not observe.  
This observation is supported by the fact that although 
variability is considered a fundamental property of blazars, 
we only see variability in about 35\% of the sample using 
the criterion of 3$\sigma$ fluctuations in both bandpasses.  Of the 
blazars for which we do see variability, 40\% fluctuate by more than 
0.4 magnitudes.  Therefore 14\% of the blazar sample is selected by 
a variability cut at 0.4 magnitudes.  In comparison, 0.7\% of the quasar 
and 0.02\% of the random samples are selected by the same cut.  
This variability criterion will clearly not include all blazars 
observed by Palomar-QUEST;  instead, it is intended to be as efficient 
as possible in selecting blazars using a method that is not affected by 
radio or X-ray biases.

\subsection{Auxiliary Cuts}

Many types of objects are known to be variable;  therefore our primary cut 
will accept more than simply AGN.  We expect to see periodic variables,
for example due to geometric effects as in eclipsing binary systems, or 
due to pulsations as in RR Lyrae.  We also expect to see aperiodic 
variables such as flaring M dwarfs or red giants.  We will select episodic 
variables as well, for example supernovae or cataclysmic variables (CVs).  
Two cuts, independent of variability amplitude, are made in order to eliminate 
a large fraction of the expected background.  The numbers of objects 
eliminated by these cuts are discussed in section \ref{results_section}.

\subsubsection{Timescale of Observation}

We do not have sufficient lightcurve information to distinguish between 
periodic variables and aperiodic ones such as AGN.  However, we can eliminate 
episodic variables by requiring the objects to have been detected over 
a span of at least 200 days.  This requirement will not be met 
by the violent explosions of supernovae and many CVs, which have timescales 
on the order of a few months or less.

\subsubsection{Galactic Latitude}

Main sequence stars in eclipsing binary systems, red giants, RR Lyrae, and 
CVs are all faint enough such that the vast majority that are bright enough 
to see are in our galaxy.  Therefore, looking at sky positions far from the 
plane of the galaxy significantly reduces our stellar contamination.  Some 
of these 
objects are also seen at high galactic latitudes;  for example, RR Lyrae are 
often studied in the galactic halo (e.g. \cite{rrlyrae2}).  For the purpose 
of finding mainly 
AGN, we restrict our search to latitudes more than 40 degrees away from 
the galactic plane.  This is a very strict cut;  there are undoubtedly 
interesting objects to be seen closer to the galactic plane.  We leave the 
examination of lower galactic latitudes until we better understand the 
current, high latitude sample. 

\subsection{Color Classification \label{color_section}}

We expect the 
variable sample to include stars from the galactic halo.  Color information 
can be used to identify common stellar types in the sample.

One approach to making color cuts could be to keep the variable objects that 
have colors similar to known blazars.  However, the optical colors of blazars 
depend on the central frequencies of the objects' SED peaks.  
Previously discovered blazars may have these frequencies biased 
by the fact that the objects were found in the radio and/or X-ray bands.  If we 
wish to select a blazar sample independent of radio and X-ray biases, we can not 
apply color cuts that are based on the colors of radio and X-ray selected blazars.  
For this reason we do not select objects based on the similarity of their colors 
to those of known blazars.  Instead, we rank objects as poor if they have 
colors similar to common stars.  It may be true that many interesting objects 
in the sample, including blazars, have colors close to stellar.  
However, by downgrading objects with stellar colors we can decrease the 
contamination of our best blazar sample and usefully rank the objects for future 
follow-up study.

To identify stellar variables we fit the Palomar-QUEST data, plus color 
information from 
other large-area surveys, to templates of typical stars.  If an 
object's colors fit well to a template then it is assumed to be a star.  
To get the widest wavelength range of data with which 
to make comparisons, we compile all available PQ UBRIriz (determined 
using the absolute calibration), SDSS ugriz 
(data release 5;  see \cite{sdss5}), GALEX fn (data release 3; see 
\cite{galex}), and 
2MASS JHK measurements (\cite{2mass}) of the variables.  The spectral 
templates published by A. J. Pickles (\cite{pickles98}), which span 
wavelengths of 1150 to 12500\AA, are convolved 
with the transmission curves of the various surveys and fitted to the 
broadband measurements.  

The ability of the color fits to identify stars depends on the amount of 
information 
available for each variable candidate.  GALEX and SDSS do not cover all 
of the PQ sky area, so some of the PQ variables have more color 
information than others.  Consequently, the color fit assigns stellar 
status more accurately for some objects than for others.  

To study this effect, sample subsets of all PQ objects, 
known quasars, known blazars, and the PQ 
variables were divided into groups based on the availability of data 
for each object.  
Different permutations of available SDSS, GALEX, and 2MASS data yield 
eight possible coverage combinations.  For each coverage combination, 
the objects were run through the template fitting, and a 
$\chi^{2}$ cut was chosen that best balanced the cut's completeness in 
selecting AGN with the purity of the resulting sample.
The number of objects tested and the percentage whose colors look nonstellar 
are given in table \ref{QB_table}.   If the color fitting perfectly reflects 
the object type, the representative sample should have a small percentage 
with non-stellar colors, while the quasars and blazars should appear entirely 
non-stellar.  The variable sample, assuming it is made up largely 
of AGN but also includes variable stars, should be somewhere in between.  
As seen in table \ref{QB_table}, the ability to distinguish the object types 
clearly varies with the amount of available color data.

\begin{table}
\begin{center}
\begin{tabular}{|l|llllll|}
\hline
\bf{Coverage} & $\chi^{2}$ \bf{Cut} & \bf{Sample} & \bf{Quasars} & \bf{Blazars} & \bf{Variables} & \bf{Cat.} \\\hline
Q + S + G + 2 & 3.0 & 23\% (1271) & 86\% (1282) & 100\% (21) & 30\% (126) & A \\
Q + S + G - 2 & 2.4 & 11\% (1817) & 86\% (7434) & 100\% (30) & 52\% (342) & B \\
Q + S - G + 2 & 2.8 & 14\% (15261) & 85\% (1388) & 100\% (107) & 29\% (438) & C \\
Q + S - G - 2 & 2.2 & 5\% (42643) & 78\% (9211) & 98\% (194) & 37\% (974) & D \\
Q - S + G + 2 & 2.0 & 69\% (1271) & 82\% (1282) & 100\% (27) & 59\% (128) & E \\
Q - S - G + 2 & 2.0 & 52\% (15261) & 75\% (1388) & 97\% (91) & 50\% (358) & F \\
Q - S + G - 2 & 2.0 & 9\% (1817) & 25\% (7434) & 92\% (36) & 28\% (577) & G \\
Q - S - G - 2 & 2.0 & 12\% (42643) & 20\% (9211) & 82\% (159) & 23\% (976) & H \\\hline
\end{tabular} 
\end{center}
\caption{Color cut statistics for different coverage combinations.  Q=Palomar-QUEST, S=SDSS, G=GALEX, 2=2MASS.  Q - S + G - 2 indicates objects for which there is PQ and GALEX data, but no SDSS or 2MASS data.  For each type of object (a random sample of PQ objects, SDSS spectroscopic quasars, known blazars, and PQ variables), what percentage pass the chi squared cut (i.e. look non-stellar)?  In parentheses is the total number of objects used in the calculation;  note the lower statistics for blazars.  Each coverage category is given a label from A to H, roughly in order corresponding to the quality of the cut.}
\label{QB_table}
\end{table}

These percentages are helpful quantities because known quasars and blazars 
are examples of highly-variable 
objects which we would like to find.  However, they are not perfectly 
indicative of the ability of the color cut to pick out interesting objects 
in the list of variables.  For example, there are more types of interesting 
objects in the sample than just UV excess selected quasars (which dominate the 
quasar sample) and radio and X-ray selected blazars (which dominate the 
blazar sample).  Furthermore, not all desired 
objects are treated equally by the cut, as shown by the fact that the 
quasar and blazar statistics give different results from each other.  
For example, the SDSS u' filter is 
critical in separating the low redshift quasars from the main sequence, 
and therefore SDSS coverage makes a very large difference in the quasar 
percentage kept by the cut.  However, SDSS coverage makes a much smaller 
difference for the blazar sample.  

To aid in the ordering of candidates for follow-up, we divide the variables 
into 8 coverage categories, A through H.  In our letter assignment we adopt 
the philosophy that more information is better, given the consideration that 
we do not know the details of how the color cuts treat all types of AGN.  
Objects in category A are covered by all four surveys, while those in category 
H are seen only in Palomar-QUEST.  
The results given in table \ref{QB_table} determine the ordering of the 
categories in cases where there is coverage by the same number of surveys, 
but different combinations (like cases B and C).  Each object is therefore 
described by its coverage (A through H) and its color classification 
(pass or fail), where the variable passes the classification if it looks 
non-stellar.

\section{Expected Source and Background Counts}

The prevalence of variable stars depends strongly on the flux limit of 
one's data.  To date there have been very few studies of the variability 
of large, heterogeneous samples of objects, therefore the characterstics 
of overall stellar 
variability, including how it scales with survey depth, are not well understood.  
The Faint Sky Variability Survey (FSVS) (\cite{fsvs}) and 
the SDSS (\cite{sesar07}) 
have both studied the variable fraction of main sequence stars.  
While their flux limits are fainter than ours, we can use their results 
as a rough estimate of the background we should expect.

The FSVS saw that 1\% of main sequence stars 
are variable, on timescales shorter than about 10 days, by at least 
0.01 magnitude.  Furthermore, they noted that 
0.07\% of the main sequence stars vary by more than 0.25 magnitudes, 
and 0.02\% vary more than 1 magnitude.  The Sloan Digital Sky Survey saw 
that 0.5\% of main sequence stars are variable, on timescales up to about 
five years, by at least 0.05 magnitude.

About 85 million objects pass the Palomar-QUEST relative calibration 
quality cuts and are 
examined for variability.  Assuming the above statistics, and assuming 
the PQ objects are dominated by main sequence stars, we should 
expect to see about 425,000 main sequence objects varying by at least 
0.05 magnitudes.  Between 
17 and 60 thousand quickly-varying main sequence stars (on timescales 
less than 10 days) should pass our variability cut of 0.4 magnitudes.  
Sparse lightcurve sampling will cause us to see sufficient variability 
in only a fraction of this number, around 15\% if our scanning 
cadence is equally suited for observing variability in stars and blazars.  
This leaves us with the expectation of seeing between 2,500 and 9,000 
main sequence variables over the entire 15,000 square degrees of the survey.  

In order to preferentially remove variable stars from our sample, 
we eliminate objects fewer than 40 degrees from the galactic plane.  
At the typical magnitude of the variables, this cut should reduce 
the stellar contribution by a factor of $\sim$13 (\cite{zombeck}).  
In our final search area, therefore, we expect to select between roughly 
190 and 700 main sequence variable stars.

The number density of blazars 
is not well known, but work on a deep sample of blazars 
(\cite{padovani07}) estimates 0.6 deg$^{-1}$ for FSRQs and 
0.06 deg$^{-1}$ for BL Lacs down to a 5 GHz flux of 50 mJy.  
It is not clear how the optical flux of these blazars relates 
to the radio flux limits used in the study.  However, if all of these 
blazars are above our optical flux limit, we would expect to find about 
10,000 over the whole survey, or 4,800 over the high galactic latitude 
area considered.  

\section{Results \label{results_section}}

Over the entire Palomar-QUEST sky area of 15,000 square degrees, 
14,185 objects pass the variability cut of 0.4 magnitudes.  
Of these, 10,838 have been seen over a timespan of at least 200 days, 
and therefore are not short-timescale transients.  
(None of the blazars in the set of 425 collected in \cite{blazar_paper} is 
removed from the sample by this cut.)  A subset of 3,955 
lie more than 40$^{\circ}$ away from the galactic plane, and so are 
less likely to be variable stars.  
The 3,955 variable candidates have been examined by eye to eliminate 
those that are clearly contaminated by artifacts or are extended galaxies 
that are likely mismeasured by the PSF photometry software.  There are 
842 apparent cases of poor data, for example if the 
lightcurves show 
a single outlier corresponding to R and I images on bad chip areas.  
This leaves 3,113 variables in the final sample.

Of the 3,113 variables, 557 have been previously identified by other surveys.  
Table \ref{known_variables_table} summarizes the identifications.  
Table \ref{known_variables_table} also gives the total number of 
identified AGN, probable AGN (including radio sources, X-ray sources, 
and color-selected quasar candidates), other known variables, and remaining 
identified objects (not known to be variable).  These IDs have been 
collected mainly using NED\footnote{http://nedwww.ipac.caltech.edu/} 
and SIMBAD\footnote{http://simbad.u-strasbg.fr/simbad/}.  

As well as collecting information from the literature, we have initiated 
a program of spectroscopic identifications of these sources at the 
Palomar 200'' telescope.  
To date a few tens of objects have been observed, with roughly a half being 
AGN, and a half being CVs and other types of variable stars.  Among the AGN, 
there is roughly an equal number of probable blazars (with featureless blue 
continuum spectra) and broad emission line quasars, most of which have radio 
source counterparts in NED and are thus likely FSRQ.  A detailed discussion 
of these observations will be presented in a forthcoming paper.

\begin{table}
\begin{center}
\begin{tabular}{|l|l|}
\hline
\bf{Object Type} & \bf{Number}\\\hline
BL Lac & 18 \\
FSRQ & 8 \\
QSO & 119 \\
QSO and Radio Source & 23 \\
QSO and Radio and Xray Source & 4 \\
QSO and Xray Source & 15 \\
Radio Source & 78 \\
Xray Source & 32 \\
Radio and Xray Source & 22 \\
UV Excess QSO Candidate & 76 \\
UV Excess QSO Candidate and Radio Source & 3 \\
Variable Object & 15 \\
RR Lyra & 38 \\
Carbon Star & 26 \\
Star & 65 \\
Galaxy & 15 \\\hline\hline
Total Identified Objects & 557 \\
Total AGN & 187 \\ 
Total Probable AGN & 211 \\
Total Other Known Variables & 79 \\
Total Other Known Objects & 80 \\\hline
\end{tabular}
\end{center}
\caption{Identification summary for known QUEST variables.}
\label{known_variables_table}
\end{table}

A measure of the success of the color classification is shown in 
table \ref{known_color_results}, which gives the color categories and 
classifications for PQ variables that are identified in the literature 
as stars and AGN.  Of known AGN, 72\% pass our color cut, i.e. have 
non-stellar colors.  Of known stars, only 17\% pass the cut.  
The table also shows that the color categories with more surveys' 
information yield more accurate results, as expected.  
The statistics are quite low, but this is a useful assessment of 
the accuracy of the classifications for our 
variable sample.  

\begin{table}
\begin{center}
\begin{tabular}{|l|l|l|l|}
\hline
\bf{Object Type} & \bf{Color Category} & \bf{\% Pass} & \bf{\# Total} \\\hline
AGN & A & 100 & 8 \\
& B & 74 & 62 \\
& C & 73 & 15 \\
& D & 75 & 75 \\
& E & 100 & 4 \\
& F & 43 & 7 \\
& G & 50 & 10 \\
& H & 33 & 6 \\
& All & 72 & 187 \\\hline
Star & A & 7 & 15 \\
& B & 19 & 16 \\
& C & 27 & 11 \\
& D & 15 & 20 \\
& E & 0 & 0 \\
& F & 0 & 0 \\
& G & 50 & 2 \\
& H & 0 & 1 \\
& All & 17 & 65 \\\hline
\end{tabular}
\end{center}
\caption{Color classifications for known AGN and stars in the variable sample.}
\label{known_color_results}
\end{table}

The Palomar-QUEST variables are available at 
http://webvoy.cacr.caltech.edu/CasJobs in the PQVariables1 table in 
the QuestProducts database.  The table's fields are described in detail 
in appendix \ref{help_appendix}.  PQVariables1 
includes, along with the position of the objects, 
the maximum 3$\sigma$ significant magnitude jump seen between any two epochs 
in both the Rr and Ii data.  The main selection
criteria for the PQ variables is that this value must be greater than 0.4.  
Also given is the RMS of the relatively calibrated Rr magnitudes of the object, 
and how many Rr measurements were available.  
Typically there are the same number of Rr and Ii measurements.  
The average Johnson R magnitude of the object, measured using the standard 
calibration, is also provided.  The color classification of the object 
is listed, as well as its coverage category, as described in section 
\ref{color_section}.  PASS means that the object {\it does not} have stellar 
colors.  As well as this information from the PQ Survey, the database 
provides available information from surveys at other wavelengths.  
Fluxes are provided in radio, infrared, UV, and X-ray wavelengths for subsets 
of the variables with such measurements in the FIRST (\cite{becker95}), 
NVSS (\cite{nvss}), CRATES (\cite{crates}), 2MASS, UKIDSS (\cite{ukidss}), 
GALEX, ROSAT (\cite{rosat}),
 or XMM (\cite{xmm}) surveys.  The fact that these objects each vary by 
more than 0.4 magnitudes in optical wavelengths complicates careful color analyses 
of these objects.  However, flux information from a large variety of wavelengths 
can give rough information about the objects' SEDs, and therefore be of interest 
when selecting certain kinds of candidates for follow-up.  
Conversely, data over a large wavelength range can provide insight into 
the physical processes of those candidates with secure identifications.  
For example, it would be interesting for the Fermi Gamma-Ray Space 
Telescope to measure the typical 
gamma-ray flux emission from blazars in this sample as opposed to objects
selected in the standard manner.

\begin{figure}
\begin{center}
\plotone{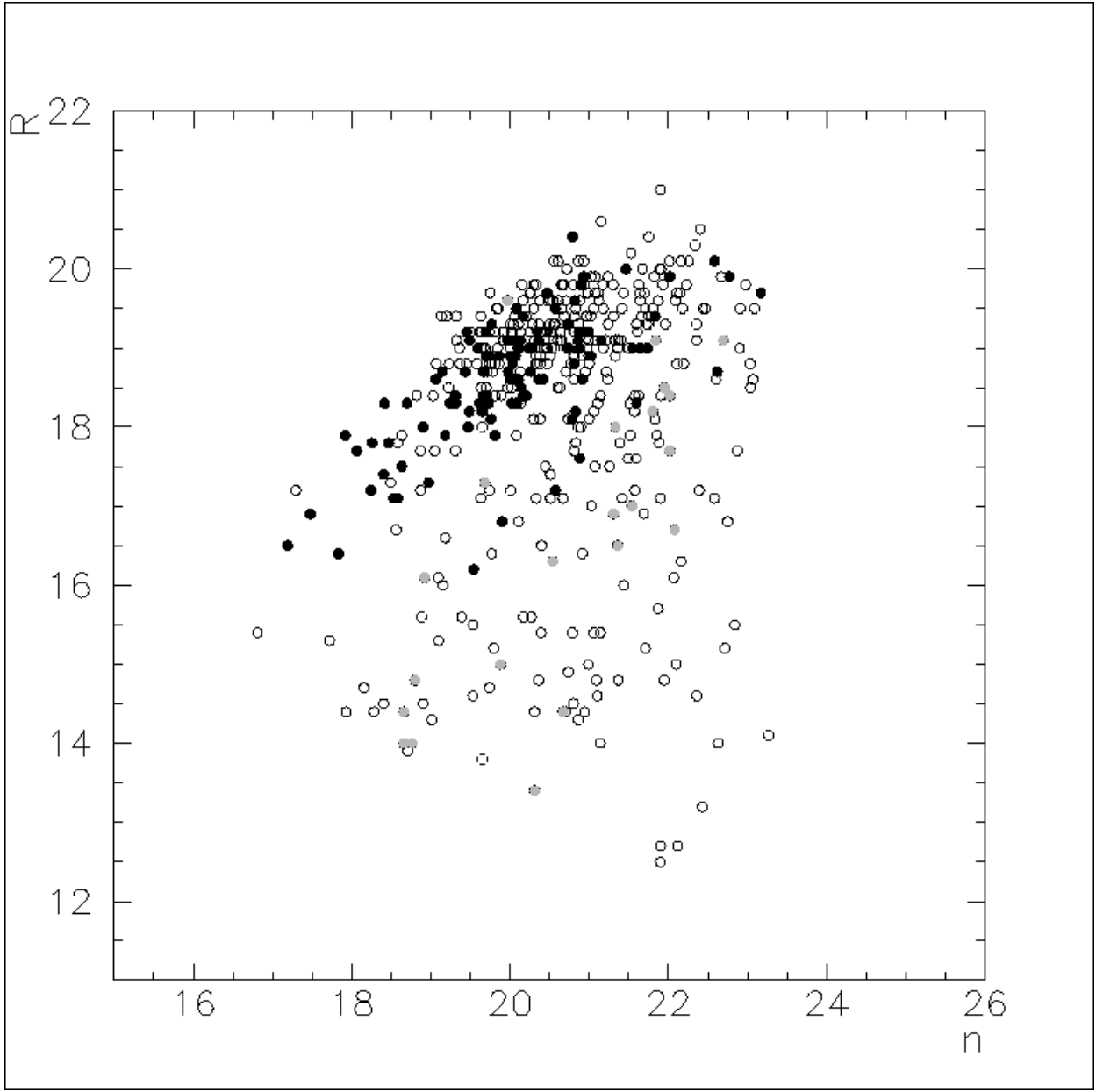}
\end{center}
\caption{Average Johnson R magnitude versus GALEX n magnitude.  Variables in color classes PASS\_A through PASS\_D with GALEX measurements are shown as hollow circles.  Known AGN are filled in black.  Known stars are filled in gray.}
\label{variables_r_n}
\end{figure}

As an example of how multiwavelength information can aid in further 
classifying the candidates, figure \ref{variables_r_n} shows the 
average Johnson R magnitude of the variables versus GALEX n magnitude, for those 
with good color classification and GALEX measurements.  The variables in 
color categories A through D with colors unlike the stellar templates 
are shown as hollow circles.  
Those objects already known to be AGN are filled with black, while those 
identified as stars are filled with gray.  The known AGN are bright in n 
while faint in R;  the majority of the variables overlap the AGN distribution.  
Known stars have similar n measurements but are brighter in R.  The bright-R 
objects are most likely not main sequence variables, however, because of their 
color classifications.  
Broad wavelength information is clearly helpful in characterizing the variables.  
For the objects with no such data, however, the PASS/FAIL color classification 
scheme summarizes the available information using just optical measurements.

\section{Conclusions}

We have examined 7,200 square degrees of relatively calibrated, 
high galactic latitude data from the Palomar-QUEST Survey and compiled a 
list of 3,113 of the most variable objects seen.  Specifically, these 
objects have 
been observed to vary by more than 0.4 magnitudes simultaneously in two 
bandpasses, over any time scale probed by the survey (from hours to 3.5 
years).  We insist that the variable candidates have been detected by the 
survey over a span of at least 200 days in order to eliminate transients 
from the sample, and we have examined the objects by eye to eliminate 
those that appear variable due to obvious data artifacts.  
To try to identify 
variable stars in the sample we have compared the objects' colors to 
spectral templates of common stellar types.  The 3,113 variables are 
thereby grouped into subclasses according to the similarity between their 
colors and those of the templates.  Although there may be blazars, or 
other interesting objects, with colors similar to common 
stars, the non-stellar color classes will the the purest samples of AGN 
and therefore perhaps the highest priority for follow-up.  
The catalog is available at http://webvoy.cacr.caltech.edu/CasJobs, 
in the PQVariables1 table in the QuestProducts database.  The contents 
of the catalog are described in Appendix \ref{help_appendix}.

The search for highly variable objects in the Palomar-QUEST Survey is 
motivated by the desire to select a large sample of blazars.  
An optical 
variability-based blazar search has not previously been carried out on 
this scale;  these rare AGN are typically found through their X-ray and/or 
radio emission.  Current blazar samples are most likely influenced by 
selection effects related to the wavelength of their discovery;  those 
discovered in radio wavelengths tend to be LBLs, with lower 
frequency emission, while X-ray 
discovered blazars tend to be HBLs.  A sample found with optical 
selection effects and without color biases may 
include objects significantly different from those currently discovered, 
both in the frequency range of the emission and in the overall SED shape.  

Although wide area sky surveys are becoming more common, the time 
domain remains relatively unexplored.  It is not understood what 
a large variability-selected sample of objects may hold.  We have 
compiled a highly variable sample with the aim of selecting blazars;  
however all of the objects have exhibited dramatic variability that 
is very rare and deserves further study.  
We hope that current and 
future surveys will observe these PQ variables in order to improve 
our understanding of the physical mechanisms behind violent variability.

\acknowledgements

We would like to thank the staff of Palomar Observatory for their 
help, and Rochelle Lauer for her work with the transfer, 
processing and storage of the data.  
We thank the Office of Science of the Department of Energy 
(grant DE-FG02-92ER40704) and the National Science Foundation 
(grants AST-0407297 and AST-0407448) for support.

{\it Facilities:} \facility{PO:1.2m}

\appendix

\section{Description of the PQVariables1 Database Table \label{help_appendix}}

The PQVariables1 table is part of the QuestProducts database at \newline
http://webvoy.cacr.caltech.edu/CasJobs.  Table \ref{column_table} 
describes the contents of PQVariables1.  
Table \ref{ref_table} explains the abbreviations used in the database table.

\begin{center}
\begin{longtable}{|l|p{10cm}|}
\caption{Description of the PQVariables1 table in the \newline QuestProducts database.}
\label{column_table} \\

\hline
{\bf Column Name} & {\bf Description} \\\hline
\endfirsthead

\multicolumn{2}{c}
{{\tablename\ \thetable{} -- continued from previous page}} \\ \hline
{\bf Column Name} & {\bf Description} \\\hline
\endhead

\hline \multicolumn{2}{|r|}{{Continued on next page...}} \\ \hline
\endfoot

\hline
\endlastfoot

ra & Right ascension in J2000 degrees \\
dec & Declination in J2000 degrees \\
delta\_mag & The maximum magnitude jump seen between any two epochs.  
To be significant, the jump must be seen in both Rr and Ii bands;  
the smaller of the two bands' magnitude change is given here. \\
R\_sigma & Rr magnitude sigma \\
n\_obs & Number of Rr observations.  The number of Ii observations is typically the same. \\
R\_mean & Mean Johnson R magnitude \\
color\_class & The results of the stellar spectral template fitting.  PASS means the object does not have stellar colors.  A through H indicate the amount of color data available for the fit, with A yielding the most reliable results. \\
notes & Information taken from other datasets. \\
redshift&  Redshift, taken from other datasets. \\
redshift\_err & Redshift error \\
FIRST & 1.4 GHz flux, in mJy, from the FIRST Survey (Becker et al., ApJ 450, 559 (1995)) \\
FIRST\_err & 1.4 GHz flux error, in mJy, from the FIRST Survey \\
NVSS & 1.4 GHz flux, in mJy, from NVSS (Condon et al. 1998) \\
CRATES\_4.8 & 4.8 GHz flux, in mJy, from the CRATES Survey (Healey et al., ApJS 171, 61 (2007)) \\
CRATES\_8.4 & 8.4 GHz flux, in mJy, from the CRATES Survey \\
2MASS\_K&  K magnitude from 2MASS (Skrutskie et al., AJ 131, 1163 (2006)) \\
2MASS\_err\_K & K magnitude error from 2MASS \\
2MASS\_H & H magnitude from 2MASS \\
2MASS\_err\_H & H magnitude error from 2MASS \\
2MASS\_J & J magnitude from 2MASS \\
2MASS\_err\_J & J magnitude error from 2MASS \\
UKIDSS\_K & K magnitude from UKIDSS (Lawrence et al., MNRAS 379, 1599 (2007)) \\
UKIDSS\_err\_K & K magnitude error from UKIDSS \\
UKIDSS\_H & H magnitude from UKIDSS \\
UKIDSS\_err\_H & H magnitude error from UKIDSS \\
UKIDSS\_J & J magnitude from UKIDSS \\
UKIDSS\_err\_J & J magnitude error from UKIDSS \\
UKIDSS\_Y & Y magnitude from UKIDSS \\
UKIDSS\_err\_Y & Y magnitude error from UKIDSS \\
GALEX\_n & n magnitude from GALEX (Martin et al., ApJ 619. L1 (2005)) \\
GALEX\_err\_n & n magnitude error from GALEX \\
GALEX\_f & f magnitude from GALEX \\
GALEX\_err\_f & f magnitude error from GALEX \\
ROSAT & 0.1-2.4 keV counts per second from ROSAT (Voges et al., A\&A 349, 389 (1999)) \\
ROSAT\_err & 0.1-2.4 keV counts per second error from ROSAT \\
XMM & 0.2-12 keV erg/s/cm\verb|^|2 flux from XMM (Jansen et al., A\&A 365, L1 (2001)) \\
XMM\_err & 0.2-12 keV erg/s/cm\verb|^|2 flux error from XMM \\
QSO & flag: Has the object been identified as a QSO?  Further information in the notes field. \\
carbon\_star & flag: Has the object been identified as a carbon star?  Further information in the notes field. \\
RR\_Lyra & flag: Has the object been identified as an RR Lyra?  Further information in the notes field. \\
BL\_Lac & flag: Has the object been identified as a BL Lac?  Further information in the notes field. \\
FSRQ & flag: Has the object been identified as an FSRQ?  Further information in the notes field. \\ \hline
\end{longtable}
\end{center}

\begin{center}
\begin{longtable}{|l|p{10cm}|}
\caption{Explanation of abbreviations in the PQVariables1 table.}
\label{ref_table} \\
\hline
{\bf Abbreviation} & {\bf Reference} \\\hline
\endfirsthead

\multicolumn{2}{c}
{{\tablename\ \thetable{} -- continued from previous page}} \\ \hline
{\bf Abbreviation} & {\bf Reference} \\\hline
\endhead

\hline \multicolumn{2}{|r|}{{Continued on next page...}} \\ \hline
\endfoot

\hline
\endlastfoot

 QUESTII & The Palomar-QUEST Survey \\
 QUEST1 & The QUEST1 Variability Survey (Rengstorf et al., ApJ 617, 184 (2004)) \\
 1 Jansky & A Complete Sample of 1 Jansky BL Lacs (Stickel et al., ApJ 374, 431S (1991)) \\
 2QZ & The 2DF QSO Redshift Survey (Croom et al., MNRAS 322, 29 (2001)) \\
 APMUKS & The APM Galaxy Survey (Maddox et al., MNRAS 243, 692 (1990)) \\
 BB93 & Beauchemin and Borra, AJ 105, 1587 (1993) \\
 BF80 & Berger and Fringant, A\&AS 39, 39 (1980) \\
 CGCS & Carbon Stars discovered by the SDSS (Margon et al., AJ 124, 1651 (2002)) \\
 EDCSN & The ESO Distant Cluster Survey (White et al., A\&A 444, 365 (2005)) \\
 DWS & A Catalog and Atlas of CVs (Downes, Webbink, and Shara, PASP 109, 345 (1997)) \\
 FASTT & Variables in the SDSS Calibration Fields (Henden and Stone, AJ 115, 296 (1998)) \\
 FBQS & FIRST Bright Quasar Catalog (White et al., ApJs 126, 133 (2000)) \\
 GAMMA1 & Gamma-Ray Blazars (Sowards-Emmerd et al., ApJ 626, 95S (2005)) \\
 HE & The Hamburg/ESO Survey (Christlieb et al., A\&A 431,143 (2005)) \\
 Kukarkin & ID List of Variables Nominated in 1968 (Kukarkin et al., IAU. Inform. Bull. Var. Stars, 311,1 (1968)) \\
 M53 & IDs of Stars in the Globular Cluster M53 (Rey et al., AJ 116, 1775 (1998)) \\
 JVAS & The Jodrell Bank VLA Astrometric Survey (Browne et al., MNRAS 293, 257 (1998)) \\
 PUL3 & The Pul-3 Catalogue of 58483 Stars in the Tycho2 system (Khrutskaya, Khovritchev, and Bronnikovan, A\&A 418, 357 (2004)) \\
 LBQS & The Large Bright Quasar Survey (Hewett et al., AJ 109, 1498 (1995)) \\
 MAPS-NGP & The MN Automated Plate Scanner Catalog of the North Galactic Pole (Cabanela et al., PASP 115, 837 (2003)) \\
 Mauron & Cool Carbon Stars in the Halo (Mauron et al., A\&A 418, 77 (2004)) \\
 Meinunger & RR Lyrae-type stars in Sonneberg fields (Meinunger, Astron. Nachr. 298, 171 (1977)) \\
 MIP & McIntosh, Impey, and Petry, AJ 128, 544 (2004)  \\
 Palomar 13 & IDs of Stars in the cluster Palomar 13 (Blecha et al., A\&A 419, 533 (2004)) \\
 PKS & Parkes Catalog (Wright, A. E., \& Otrupcek, R. E. 1990, PKSCAT90: Radio Source Catalogue and Sky Atlas) \\
 ROSAT & The Roentgen Satellite (Voges et al., A\&A 349, 389 (1999)) \\
 Siegel & RR Lyrae stars in Bootes (Siegel, ApJ 649, L83 (2006)) \\
 SDSS & The Sloan Digital Sky Survey (Adelman-McCarthy et al., ApJS 175, 297 (2008)) \\
 SDSS & Candidate RR Lyrae found in the SDSS commissioning data (Ivezic et al., AJ 120, 963 (2000)) \\
 VCV & The Veron-Cetty \& Veron Catalogue of Quasars and Active Nuclei (Veron-Cetty and Veron, A\&A 455, 773 (2006)) \\
 Vivas & The QUEST RR Lyrae survey (Vivas et al., AJ 127, 1158 (2004)) \\
\hline
\end{longtable}
\end{center}

\end{document}